\documentclass[useAMS,usenatbib,usegraphicx,letters]{mn2e}

\usepackage{graphicx,times}

\title[The $^{13}$Carbon footprint of \protect{B[e]} supergiants]{The
$\bf{^{13}}$Carbon footprint of B[e] supergiants\thanks{Based on 
observations collected with the ESO VLT Paranal Observatory under 
programme 384.D-1078(A)}}
\author[A. Liermann et al.]{A.~Liermann,$^{1,2}$\thanks{E-mail: 
liermann@mpifr-bonn.mpg.de; kraus@sunstel.asu.cas.cz; 
oschnurr@aip.de; borges@on.br} 
M.~Kraus,$^{3}$\footnotemark[2] O.~Schnurr$^{4,5}$\footnotemark[2] and 
M.~Borges Fernandes$^{6}$\footnotemark[2]\\
$^{1}$ Universit\"at Potsdam, Institut f\"ur Physik und Astronomie,
Karl-Liebknecht-Str. 24-25, 14476 Potsdam, Germany\\
$^{2}$ Max-Planck-Institut f\"ur Radioastronomie, Auf dem H\"ugel 69,
53121 Bonn, Germany\\
$^{3}$ Astronomick\'y \'ustav, Akademie v\v{e}d \v{C}esk\'e republiky,
Fri\v{c}ova 298, 251\,65 Ond\v{r}ejov, Czech Republic\\ 
$^{4}$ University of Sheffield, Department of Physics and Astronomy, University of Sheffield, Sheffield S3 7RH, UK\\
$^{5}$ Astrophysikalisches Institut Potsdam, An der Sternwarte 16, 14482
Potsdam, Germany\\
$^{6}$ Observat\'orio Nacional, Rua General Jos\'e Cristino 77,
20921-400 S\~ao Cristov\~ao, Rio de Janeiro, Brazil
}

\begin{document}

\date{Accepted. Received; in original form}

\pagerange{\pageref{firstpage}--\pageref{lastpage}} \pubyear{2010}

\maketitle

\label{firstpage}

\begin{abstract}
  We report on the first detection of $^{13}$C enhancement in two B[e]
  supergiants in the Large Magellanic Cloud. Stellar evolution models
  predict the surface abundance in $^{13}$C to strongly increase
  during main-sequence and post-main sequence evolution of massive
  stars. However, direct identification of chemically processed
  material on the surface of B[e] supergiants is hampered by their
  dense, disk-forming winds, hiding the stars. Recent theoretical
  computations predict the detectability of enhanced $^{13}$C via the
  molecular emission in $^{13}$CO arising in the circumstellar disks
  of B[e] supergiants. To test this potential method and to
  unambiguously identify a post-main sequence B[e]SG by its $^{13}$CO
  emission, we have obtained high-quality $K$-band spectra of two
  known B[e] supergiants in the Large Magellanic Cloud, using the Very
  Large Telescope's Spectrograph for INtegral Field Observation in the
  Near-Infrared (VLT/SINFONI). Both stars clearly show the $^{13}$CO
  band emission, whose strength implies a strong enhancement of
  $^{13}$C, in agreement with theoretical predictions. This first
  ever direct confirmation of the evolved nature of B[e] supergiants
  thus paves the way to the first identification of a Galactic B[e]
  supergiant.

\end{abstract}


\section{Introduction}

B[e] supergiants (B[e]SGs) are massive and luminous B stars with
strong, non-spherical, disk-forming winds. The conditions in terms of density
and temperature within these disks are ideal for efficient molecule and dust 
formation, traced by CO band emission \citep{McGregor, Morris} and a 
strong near and mid-infrared excess \citep{Zickgraf86}.
However, the usual distance issue renders it 
difficult to obtain useful luminosity estimates to classify a Galactic B[e]
star definitely as B[e]SG \citep[e.g.,][]{Lamers, Kraus09}. An additional 
complication arises from the position of Galactic B[e]SG candidates in the 
empirical Hertzsprung-Russell diagram, which they share with the
pre-main sequence Herbig Ae/Be  
stars. Even worse, the latter are also surrounded by gas and dust disks
\citep{Waters, WatersWaelkens}, giving rise to identical observable features 
such as the infrared excess and the well-pronounced CO band emission 
\citep{BikThi, Thi, Bik}. For an unambiguous classification of a Galactic
B[e] star as B[e]SG, a reliable tracer based on chemical processing during  
the evolution of a massive star is thus needed.

   \begin{table*}
      \caption{Parameters of the observed B[e]SGs.}
      \label{tab:parameters}
     \begin{center}
         \begin{tabular}{@{}lccccccccc}
            \hline
Object & Sp.\,Type &$T_{\rm eff}$ & $L_{*}$ & $R_{*}$ & $M_{\rm ini}$ & $E(B-V)$ & $v_{\rm sys}$ & inclination & Reference \\
LHA    &  & [$10^{3}$\,K] & [$10^{5}\,L_{\sun}$] & [$R_{\sun}$] & [$M_{\sun}$] & & [km\,s$^{-1}$] &  &  \\
            \hline
120-S 12 & B1\,I & 23 & 2.2 & 30  & $\sim 25$ & 0.25 & $285\pm 10$ & intermediate & \citet{Zickgraf86} \\
120-S 73 & B8\,I & 12 & 3.0 & 125 & 25-30     & 0.12 & $264\pm 3$ & $\sim$ pole-on & \citet{Stahl} \\
            \hline
         \end{tabular}
     \end{center}
   \end{table*}
                                                                                
Based on stellar evolution models for Galactic stars \citep{Schaller,
MeynetMaeder03}, \citet{Kraus09} recently pointed out the significant
enrichment of the stellar surface abundance in $^{13}$C already during
the main-sequence evolution of massive stars. However, while the
surface of B[e]SGs is usually hidden within their dense winds and
disks, \citet{Kraus09} proposed 
that this enrichment should be detectable in the ejected material via strongly 
enhanced $^{13}$CO band emission from the disk-forming winds of B[e]SGs.
If true, we would have, for the first time, an easy and robust method to
unambiguously distinguish between pre- and post-main sequence evolution and to 
classify Galactic B[e] stars as supergiants, because Herbig Ae/Be stars do not 
show any $^{13}$C enhancement. However, before this method can be applied to 
the Galactic B[e] stars, its validity needs to be tested on well-known 
B[e]SGs. The best-known sample of confirmed B[e]SGs is located in the 
Magellanic Clouds \citep{Lamers, Zickgraf06}.

To test the $^{13}$CO method, we have observed two B[e]SGs in the Large 
Magellanic Cloud (LMC) which are known to display $^{12}$CO band emission 
originating in their circumstellar disks. In this Letter we report on the 
first definite detection of $^{13}$C enrichment from B[e]SGs via 
enhanced $^{13}$CO band emission whose presence is direct evidence for the 
evolved nature of our programme stars. 
 
\begin{figure}
 \includegraphics[width=\columnwidth]{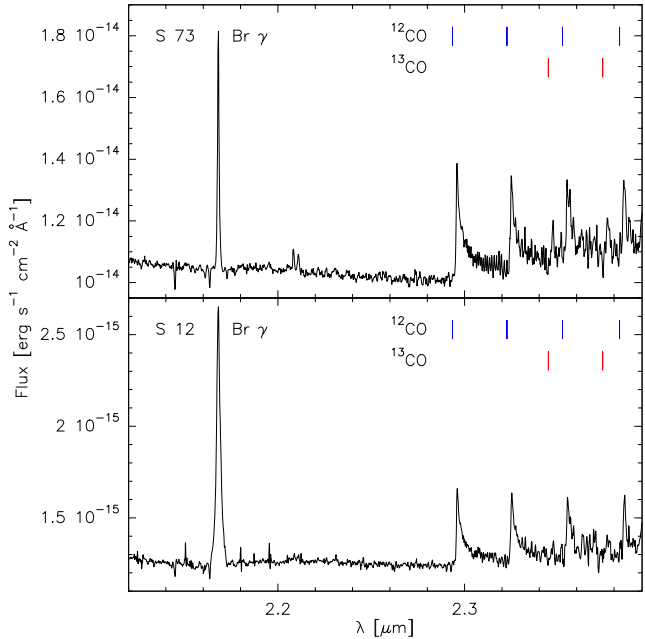}
\caption{Parts of the observed, flux-calibrated SINFONI $K$-band spectra of the 
B[e]SG stars S\,73 (top) and S\,12 (bottom) covering the Br$\gamma$ line and 
the first-overtone bands in both $^{12}$CO and $^{13}$CO. The wavelengths of 
the band heads are indicated by vertical ticks.}
  \label{fig:spectra}
\end{figure}

\section[]{Observation and reduction}
\label{sec:obs}
Between October and November 2009, we have obtained high-quality $K$-band 
spectra 
(1.95 - 2.45\,$\mu$m) of the two LMC B[e]SG stars LHA\,120-S\,12 and 
LHA\,120-S\,73 (in the following S\,12 and S\,73 respectively), using the
Spectrograph for INtegral Field Observation in the Near-Infrared (SINFONI)
\citep{Eisenhauer+2003, Bonnet+2004}. 
Observations were carried out in an AB pattern with a field of view of
8\arcsec $\times$ 8\arcsec. 
B-type standard stars for flux calibration were observed at similar airmass.

\begin{figure}
 \includegraphics[width=\columnwidth]{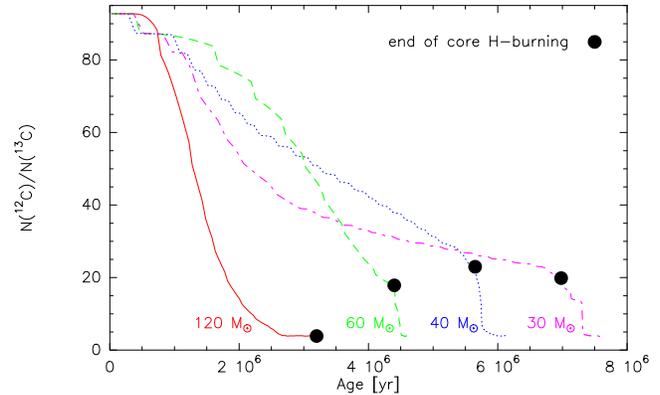}
  \caption{Change in $^{12}$C/$^{13}$C surface abundance ratio
     during the evolution of rotating massive stars at LMC metallicity, based
     on the models of \citet{MeynetMaeder05}.}
  \label{fig:ratio}
\end{figure}

Data reduction was performed with the SINFONI pipeline 
(version 2.0.5). Raw frames were corrected for bad
pixels, distortion, and flatfield, followed by the wavelength
calibration. 
For flux calibration, we extracted the standard-star spectra and
scaled them with the according Kurucz model \citep{Kurucz} to their 2MASS
$K_{\rm s}$-band magnitude \citep{Skrutskie} to obtain calibration curves.
The final flux-calibrated spectra have a signal-to-noise ratio $S/N \sim 500$
and a spectral resolution $R \sim 4500$ at $2.3\,\mu$m. 
Figure\,\ref{fig:spectra} shows parts of the $K$-band spectra 
covering the Br$\gamma$ line of hydrogen and the CO bands.

\section{Analysis and Results}

The two LMC B[e]SGs listed in Table\,\ref{tab:parameters} 
were selected because they are known to display strong $^{12}$CO band
emission \citep{McGregor}. The pole-on or intermediate orientation of 
their disks guarantees that we see the complete CO emitting disk area. The
spectral range of SINFONI covers the first four band heads of $^{12}$CO 
and the first two of $^{13}$CO. Visual inspection of the spectra 
(Fig.\,\ref{fig:spectra}) reveals that the $^{12}$CO band heads are 
prominent in both stars; while $^{13}$CO is clearly visible in S\,73 (top 
panel), its presence in S\,12 is slightly less obvious. 

\begin{figure*}
 \includegraphics[width=\textwidth]{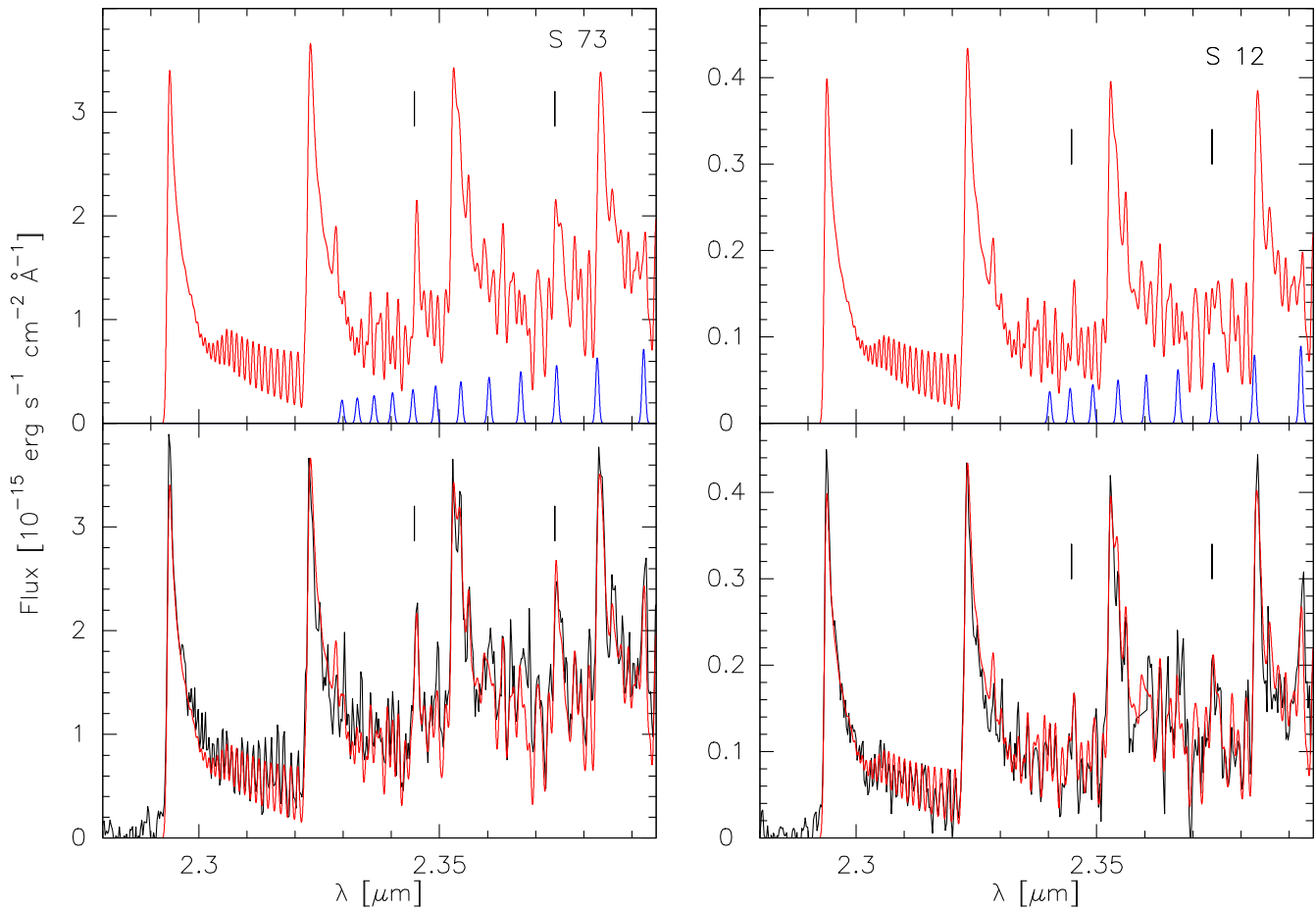}
  \caption{{\it Top:} Synthetic spectra of the CO bands (red) and of the 
Pfund series (blue). {\it Bottom:} Combined CO band plus Pfund spectra (red)
overlayed on the observed (black), flux-calibrated CO band spectra of S\,73 
(left) and S\,12 (right). The vertical ticks in the respective panels indicate 
the positions of the $^{13}$CO band heads.}
 \label{fig:fits}
\end{figure*}

We corrected the spectra for the systemic velocity and dereddened them with the
$E(B-V)$ values from Table\,\ref{tab:parameters} applying the interstellar
extinction curve of \citet{Howarth} and using $A_{V}/E(B-V) = 3.1$. 
To obtain a flux-calibrated, quasi-pure CO emission spectrum, we 
fitted and subtracted a linear pseudo-continuum longwards of $2.25\,\mu$m.
The spectra of the two stars do not show indication for significant
emission of the hydrogen Pfund series, but some minor contribution cannot
be excluded and might have to be taken into account later on in the analysis.

Inspection of the CO spectra reveals that the band heads of $^{12}$CO all have
about equal strength, which means that, surprisingly, the CO gas is rather 
cool ($T_{\rm CO} < 3000$\,K). We will come to this aspect again in 
Sect.\,\ref{sec:discuss}. On the other hand, the excitation of pronounced band 
head structures requires CO temperatures in excess of $\sim 2000$\,K. 
Therefore, the location of the CO emitting region must be closer to the star 
than the dusty regions, because dust sublimates at $\sim 1600$\,K
\citep[e.g.,][]{Lodders}. Moreover, the rise of the CO 
quasi-continuum towards longer wavelengths indicates substantial optical 
thickness of the emission. To model the CO bands we can thus assume that the 
emission is in local thermodynamical equilibrium (LTE) and, preempting results 
from Sect.\,\ref{sec:discuss}, that the CO gas is located in a ring around the 
star which has constant temperature and column density. These assumptions are 
reasonable since the hottest CO component, which arises close to the star where 
usually the disk density is highest, dominates the total CO spectrum 
completely. For details on the CO band modeling we refer to \citet{Kraus00}
and \citet{Kraus09}.

To constrain the range of the $^{12}$C/$^{13}$C abundance ratio in the disk, we
took expected $^{12}$C/$^{13}$C surface abundance ratios from the stellar
evolution models of \citet{MeynetMaeder05}
for massive stars with LMC metallicity and initial rotation velocities of
300\,km\,s$^{-1}$. Fig.\,\ref{fig:ratio} shows how this ratio decreases with
time for stars of different initial mass.
The stars S\,12 and S\,73 have evolved from progenitors
with $M_{\rm ini} = 25\ldots 30\,M_{\sun}$ (see Table\,\ref{tab:parameters})
whose surface abundance should have dropped to a value
$^{12}$C/$^{13}$C $\la 20$ at the turn-off from the main sequence and decreases
down to a value of $\sim 4$ at late evolutionary phases.
We computed synthetic CO band spectra for temperatures in the range 2000 
-- 3000\,K, $^{12}$C/$^{13}$C ratios varying from 4 to 20, and CO column
densities higher than $10^{21}$\,cm$^{-2}$ to account for the optical
depth effects. We did not find any indication for additional line 
broadening from either Keplerian rotation or outflow signatures, meaning that 
the CO gas has no 
significant line-of-sight velocity component, in agreement with an (almost) 
pole-on orientation of the disks. Therefore, we used the thermal velocity as 
the intrinsic gas velocity and convolved the resulting spectrum with the 
resolution of SINFONI. Pure CO models that match the observations best
are shown in red in the top panels of Fig.\,\ref{fig:fits}, with the 
parameters listed in Table\,\ref{tab:bestfit}.

The presence of some additional emission features, especially at the red end of 
the spectrum, can be explained by the (if weak) recombination lines of the 
hydrogen Pfund series. Following the description of \citet{Kraus00}, we modeled 
the Pfund series assuming a constant electron density and an 
electron temperature of $T_{e} = 10^{4}$\,K. The lack of Pfund lines higher 
than Pf\,34 (S\,73) and Pf\,31 (S\,12) means that due to pressure ionization in 
a high-density gas, recombination into such high-energy states is suppressed. 
This is in good agreement with the electron densities of $n_{e} \ga 
10^{9}$\,cm$^{-3}$ needed for the lines to be formed in LTE conditions, as is 
obvious from their intensity ratios. As in the case of the CO bands, no 
additional line broadening is required to match the width of the Pfund lines. 
This could indicate that the Pfund emission is generated either in the disk but 
much closer to the star than the CO bands, or in the polar wind regions so 
close to the star that the wind velocity is still very small compared to the 
velocity resolution of SINFONI. The resulting synthetic Pfund series are 
included as the blue lines in the top panels of Fig.\,\ref{fig:fits}, and the
combined model spectra of CO bands and Pfund lines are overplotted in red
on the observed spectra in black in the bottom panels.

\begin{table}
 \caption{Best-fit parameters for the CO band emission from the disks of S\,12 
          and S\,73.}
  \label{tab:bestfit}
     \begin{center}
       \begin{tabular}{@{}lcccc}
          \hline
Object & $T_{\rm CO}$ & $N_{\rm CO}$ & $^{12}$C/$^{13}$C & $A_{\rm CO} \cos i$ \\
LHA    & [K] & [$10^{21}\,$cm$^{-2}$] &  & [AU$^{2}$]  \\
            \hline
120-S 12 & $2800\pm 500$ & $2.5\pm 0.5$  & $20\pm 2$ & $2.58\pm 0.15$ \\
120-S 73 & $2800\pm 500$ & $3.5\pm 0.5$  & $9\pm 1$ & $21.0\pm 0.3$ \\
            \hline
         \end{tabular}
 \end{center}
\end{table}

\section{Discussion}
\label{sec:discuss}

Our analysis has shown that the spectra of the B[e]SGs S\,73 and S\,12 display
emission from the $^{13}$CO band heads (see Fig.\,\ref{fig:fits}) which means
that their circumstellar disk material is strongly enriched with $^{13}$C;
hence, these stars are indeed evolved post-main sequence objects.

Our model computations have also confirmed the relatively low temperature of
the CO gas as indicated by the about equal strengths of the $^{12}$CO band
heads. Such low temperatures are not expected within the canonical picture of
B[e]SGs, which assumes that their disks are formed by enhanced mass flux from
the stars' equatorial regions. If the disks of the two B[e]SGs studied here
were continuously supplied, then the CO in the gaseous parts of the disk
should be visible from the hottest component having temperatures close to the
CO dissociation temperature of 5000\,K. In this case the observable CO band
spectrum, which always reflects the hottest available CO component, would be
dominated by the peaks of the higher vibrational transitions occurring at
longer wavelengths, while the lowest band head (at 2.3\,$\mu$m) would be
strongly suppressed \citep[see][]{Kraus09}. The apparent deficiency in CO gas
hotter that $\sim 2800$\,K therefore suggests that there is no CO gas closer
to the star. Hence, the disks seen around S\,12 and S\,73 cannot extend down
to the stellar surface. Instead it would be more appropriate to assume that
the stars are surrounded by a dense and cool {\it ring} of
material. Support for such a scenario comes from the recent detection of a
detached ring in quasi-Keplerian rotation around the Small Magellanic
Cloud (SMC) B[e]SG star LHA\,115-S\,65 \citep{Kraus2010}.

Alternatively, gravitational darkening according to the von Zeipel theorem
\citep{vonZeipel} in rapidly rotating stars can lead to equatorial
temperatures being sufficiently low (i.e., $\la 3000$\,K) for the hottest CO
gas to not exceed $\sim 2800$\,K. In this picture, the disk still could
extend from the stellar surface out to large distances, with the CO emitting
gas located closest to the star within the innermost, hottest disk
region. While this scenario is somewhat speculative (and heating of the disk
through light from hotter regions of the star would have to be
considered as well), we can estimate the minimum rotation speed of the
stars required to achieve an equatorial surface temperature of 3000\,K.

\begin{figure}
 \includegraphics[width=\columnwidth]{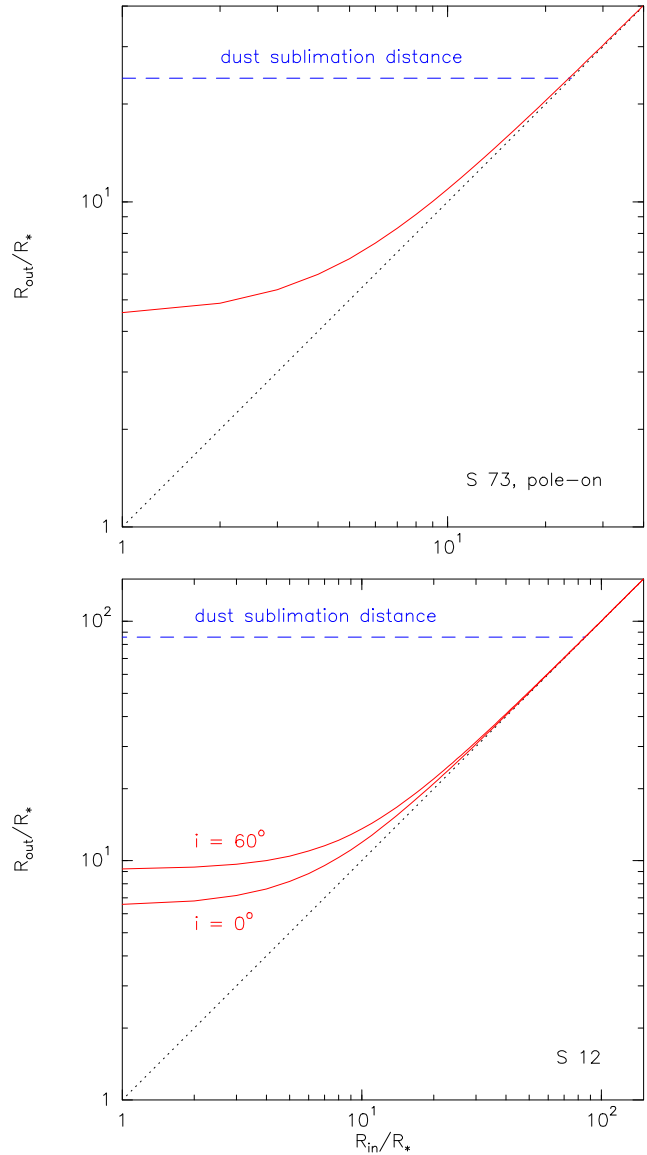}
  \caption{Outer radius of the CO emitting ring as a function of inner radius
(solid lines) for S\,73 (top) and S\,12 (bottom). The maximum outer radius is
set in both cases by the distance for dust sublimation (dashed lines). For
S\,12 two inclination angles for the disk are considered.}
  \label{fig:rout}
\end{figure}

Stellar rotation leads the poles to be hottest. Assuming that both stars are
seen pole-on and that their effective temperatures, $T_{\rm eff}$, listed in
Table\,\ref{tab:parameters} correspond to the polar values, requires a
decrease in $T_{\rm eff}$ from polar to equatorial values by factors of 7.7
(S\,12) and 4 (S\,73). This, in turn, requires the stars to rotate with $\gg
99\%$ of their critical velocity \citep[see Fig.\,5 of][]{Kraus06}. The
assumed pole-on orientation only provides us with {\it lower limits} of
the real rotation velocity, but even in the unexpected case that both stars
are seen perfectly pole-on, rotation rates close to or even at the critical
limit are highly unlikely. For comparison, the two most rapidly rotating
B[e]SGs known, the SMC stars LHA\,115-S\,23 and
LHA\,115-S\,65, rotate with $\ga 75\%$ of their critical velocity
\citep{Zickgraf00, Kraus08, Kraus2010}. 
If our program stars rotated at $\sim 75\%$ of their critical
velocity, resulting equatorial temperatures would still be $\sim 72\%$ of the
polar values (cf. Table\,\ref{tab:parameters}), i.e. much hotter than
3000\,K. The deficiency of CO gas hotter than 2800\,K is thus real, and
the most plausible scenario is hence that the molecular gas
is located within a detached ring of material rather than in a 
disk extending down to the stellar surface.

In Table\,\ref{tab:bestfit} we include the sizes of the CO emitting regions
(projected to the line of sight, $A_{\rm CO} \cos{i}$) as obtained
from our best-matching models. 
From these sizes we estimate the radial extension of the emitting CO rings.
We compute the outer radius of the ring as a function of its inner radius.
The results for both stars are depicted by the solid lines in 
Fig.\,\ref{fig:rout} indicating that for increasing inner radii the emission 
area converges towards infinitesimally narrow rings.
With the stellar luminosities given in Table\,\ref{tab:parameters} and an
assumed dust sublimation temperature of $T_{\rm sub}\simeq 1600$\,K
\citep[e.g.,][]{Lodders} we find individual sublimation radii of $R_{\rm
sub} \simeq 24\,R_{*}$ and $\simeq 86\,R_{*}$ for S\,73 and S\,12,
respectively. These maximum outer edges of the CO rings are shown in
Fig.\,\ref{fig:rout} as dashed lines. An estimate for the inner edge of the CO
ring is less precise. The CO gas has to have a temperature of about
2800\,K, only slightly higher than the dust sublimation temperature
but definitely much lower than the effective temperatures of the stars. A
reasonable estimate could thus be $R_{\rm in} \ga 10\,R_{*}$ for S\,73 and
(much) larger for S\,12. For the latter we plotted the relation between inner
and outer radius for two different values of disk
inclination. However, independent
of the exact value of $R_{\rm in}$ (and $i$ as in the case of S\,12) we can
conclude that the CO emitting area is concentrated within a rather narrow ring
at short distance from the central star.

\section{Conclusions}

Using high-quality $K$-band spectra obtained with SINFONI, we detected,
for the first time, a 13C enhancement in two B[e]SGs in the LMC,
thereby confirming their evolved nature. 
The enrichment of their disk-forming wind material with $^{13}$C and
the resulting strong $^{13}$CO band emission, as was suggested by
\citet{Kraus09}, is thus shown to be a robust method for the
identification of evolved stars. 
This method can now be applied to the Galactic B[e]SG candidates to 
disentangle the real Galactic B[e]SGs from the more abundant Herbig Ae/Be stars.

In our analysis, we also found a deficiency of hot CO gas in the disks
of both stars. This can be explained in two ways. Either both stars
rotate at their critical limit so that equatorial surface temperatures
(and hence, maximum disk temperatures) drop below $\sim 3000$\,K and
no hot CO gas is expected, or the CO emitting region is located in a
high-density, detached ring of material rather than in an equatorially
enhanced, disk-forming wind. While the former scenario cannot be
excluded, we consider the latter as the more realistic one since it
also agrees with the recently found detached gas ring around the SMC
B[e]SG star LHA\,115-S\,65 by \citet{Kraus2010}.

\section*{Acknowledgments}

A.L. is supported by the Max-Planck-Institut f\"ur Radioastronomie,
M.K. acknowledges financial support from GA \v{C}R grant number
205/08/0003. Financial support was provided to O.S. by the Science and 
Technology Facilities Council (UK). M.B.F. acknowledges Conselho Nacional de
Desenvolvimento Cient\'{i}fico e Tecnol\'{o}gico (CNPq-Brazil) for the
post-doctoral grant.

\bsp
\label{lastpage}
\end{document}